# Role of Upwelling on Larval Dispersal and Productivity of Gooseneck Barnacle Populations in the Cantabrian Sea: Management Implications

**Antonella Rivera[1]\*, Nicolás Weidberg[1]¤, Antonio F. Pardiñas[1], Ricardo González-Gil[1], Lucía García-Flórez[2], J. L. Acuña[1]**

1 Departamento de Biología de Organismos y Sistemas, Universidad de Oviedo, Oviedo, Spain, 2 Centro de Experimentación Pesquera. Consejería de Agroganadería y Recursos Autóctonos, Gijón, Spain

## Abstract

The effect of coastal upwelling on the recruitment and connectivity of coastal marine populations has rarely been characterized to a level of detail to be included into sound fishery management strategies. The gooseneck barnacle (*Pollicipes pollicipes*) fishery at the Cantabrian Coast (Northern Spain) is located at the fringes of the NW Spanish Upwelling system. This fishery is being co-managed through a fine-scale, interspersed set of protected rocks where each rock receives a distinct level of protection. Such interspersion is potentially beneficial, but the extent to which such spacing is consistent with mean larval dispersal distances is as yet unknown. We have simulated the spread of gooseneck barnacle larvae in the Central Cantabrian Coast using a high-resolution time-series of current profiles measured at a nearshore location. During a year of high upwelling activity (2009), theoretical recruitment success was 94% with peak recruitment predicted 56 km west of the emission point. However, for a year of low upwelling activity (2011) theoretical recruitment success dropped to 15.4% and peak recruitment was expected 13 km east of the emission point. This is consistent with a positive correlation between catch rates and the Integrated Upwelling Index, using a 4-year lag to allow recruits to reach commercial size. Furthermore, a net long-term westward larval transport was estimated by means of mitochondrial cytochrome c oxidase subunit I (COI) sequences for five populations in the Cantabrian Sea. Our results call into question the role of long distance dispersal, driven by the mesoscale processes in the area, in gooseneck barnacle populations and point to the prevalent role of small-scale, asymmetric connectivity more consistent with the typical scale of the co-management process in this fishery.

**Citation:** Rivera A, Weidberg N, Pardiñas AF, González-Gil R, García-Flórez L, et al. (2013) Role of Upwelling on Larval Dispersal and Productivity of Gooseneck Barnacle Populations in the Cantabrian Sea: Management Implications. PLoS ONE 8(11): e78482. doi:10.1371/journal.pone.0078482

**Editor:** Hans G. Dam, University of Connecticut, United States of America

**Received** July 5, 2013; **Accepted** September 12, 2013; **Published** November 13, 2013

**Copyright:** © 2013 Rivera et al. This is an open-access article distributed under the terms of the Creative Commons Attribution License, which permits unrestricted use, distribution, and reproduction in any medium, provided the original author and source are credited.

**Funding:** This work was financed by the Spanish Government through projects COSTAS (CTM2006-05588/MAR, Ministry of Education and Science) and DOSMARES (CTM2010-21810-C03-02, Ministry of Science and Innovation), and by the Principality of Asturias Government through project FRENTES (IB08-122, FICYT). AR and RGG were supported by a FPU fellowship (ref. AP2010-5376 and AP2008-03992, respectively, Ministerio de Educación de España) and AFP by a "Severo Ochoa" FICYT-PCTI fellowship (ref. BP09038, Asturias Regional Government). The funders had no role in study design, data collection and analysis, decision to publish, or preparation of the manuscript.

**Competing Interests:** The authors have declared that no competing interests exist.

\* E-mail: riveralidia@uniovi.es

¤ Current address: Coastal Research Group, Department of Zoology and Entomology, Rhodes University, Grahamstown, South Africa

## Introduction

Most marine benthic species have a pelagic dispersal stage which is essential for their persistence. Dispersal and further recruitment to the benthic habitat allows for connectivity among disjunct populations, leading to metapopulations which are globally viable in spite of the possibility of local extinctions [1,2]. Thus, connectivity is a key factor for population persistence, which explains the rapid expansion of research on larval dispersal and connectivity in marine populations [3]. Much of this research has been focused on large and mesoscale (100s of km) connectivity [4,5], but less on scales of a few kilometers [6,7]. Mesoscale connectivity is consistent with the typical management scale of finfish populations [8]. However, for coastal benthic organisms, the concept of connectivity includes post-dispersal processes (such as settlement, survival of early stages and reproduction) which take place at the shore and could be totally decoupled from those affecting larvae in the pelagic realm. In fact, many benthic populations are generally managed at much smaller scales and do not adapt well to finfish management tools [9]. For example, an emerging trend in the conservation of benthic resources incorporates co-management practices involving exclusive Territorial User Rights for Fishing (TURFs) in exchange for shared responsibility on resource management [10,11]. This kind of management practices favor the incorporation of local ecological knowledge into the regulations, which very often comprises aspects of the spatial distribution of the resource and of the fishermen's activity spanning a fine scale from a few meters to several kilometers. But, does the management scale in these TURF systems have the potential to interact with the connectivity patterns of the resource? And, are the dispersal scales of benthic resources consistent with the scale of co-management practices?

The gooseneck barnacle (*Pollicipes pollicipes*) fishery in the Cantabrian Sea offers a good opportunity to test these ideas. The fishery is located at the fringe of the NW Spanish Upwelling system, where frequent summer northeast winds cause the westward movement of surface waters and the vertical advection of deep water towards the coast [12,13]. This economically important fishery is currently being managed across all the Atlantic Iberian Coast, including Portugal, Galicia (NW coast of Spain) and the Northern Spanish Cantabrian Coast which comprises Asturias, Cantabria and the Basque Country [10,14,15,16]. In the Asturian coast (Cantabrian Sea), the resource is being co-managed at a remarkably fine-scale, with regulations affecting the status of single rocks as small as 30 m (Fig. 1). The co-management regime requires that the fishermen keep daily records of barnacle landings within each rock, offering an invaluable resource for research.

Apart from direct fishery data, a previous analysis of sequences of Cytochrome Oxidase subunit I of the mitochondrial DNA (COI) suggested that gene flow within Cantabrian gooseneck barnacle populations is governed by mesoscale hydrographic processes [17]. According to these authors, an eastwards net larval flow should be associated to the existence of the Iberian Poleward Current (IPC), a high salinity filament which flows from south to north along the slope of the Portuguese, Spanish and French shelves. Later work on the genetic structure and phylogeography of *P. pollicipes*, using a finer sampling scheme, confirmed that NW Atlantic populations are highly connected, pointing to local drift events and isolation-by-distance as the main causes behind the population structure [18].

A different approximation to the question of population connectivity has also been undertaken by the construction of biophysical models, where larval dispersion is simulated in a measured or reconstructed oceanographic flow field [19,20]. With some level of simplification, these models have been used to extract major patterns of dispersal among populations of marine organisms (see [21] for a review). However, these models have yet to be implemented in the Cantabrian Sea.

In this paper, we address the potential effects of Ekman transport on larval dispersal and productivity of gooseneck barnacle populations. First, we have used data collected with a moored current meter placed at the Central Cantabrian Coast to simulate the dispersal of *P. pollicipes* larvae. The simulations have been done for one summer of high (2009) and one summer of low (2011) upwelling activity, assuming the most likely stage-specific vertical distribution of the larvae. This procedure provided insight into both alongshore and cross-shore flow components associated with Ekman transport. Additionally, the predicted cross-shore component, responsible for larval recruitment to the coast, was estimated using catch rate records collected by the fishermen. Finally, the alongshore component, responsible for population connectivity along the coast, was inferred by estimation of the population migration rates according to previously published COI sequences. Our results indicate a small-scale, asymmetric connectivity in gooseneck barnacle populations, which matches the current co-management scale in the area.

## Materials and Methods

### Biological background for the larval dispersal model

We have simulated the dispersal of *P. pollicipes* larvae following the advection-diffusion model by White et al. [20] and Siegel et al. [19]. In essence, they modeled the trajectory of the larvae as embedded in a 2D horizontal flow field. We have modified this approach by including a surface and a deep layer separated by the thermocline, because those layers may experience contrasting water circulation associated to coastal upwelling [22] and because there is evidence of ontogenetic vertical migration behavior in barnacle larvae [23].

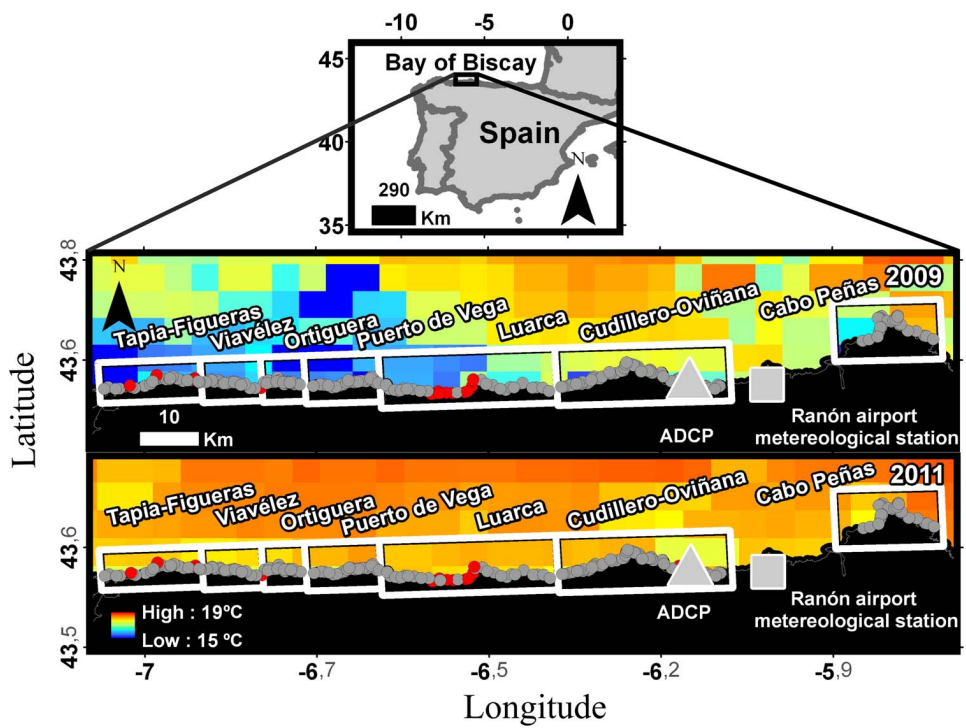

**Figure 1. Map of the Asturian gooseneck barnacle management plan.** Frames represent the seven fishers' guilds. Grey dots indicate fishing areas with no bans and red dots areas with a total ban. The light grey square indicates the meteorological station and the light grey triangle the location of the ADCP. Background colors indicate the 2009 (upper map) and 2011 (lower map) Sea Surface Temperature (SST) averaged for the period incorporated in our model (see methods). SST data were obtained from the Terra MODIS satellite at a 4 km resolution.
doi:10.1371/journal.pone.0078482.g001

Gooseneck barnacle reproduction is asynchronous [16,24]; each individual may brood up to 4 times per spawning season [24]. For the model, we have assumed a uniform release of larvae from the 1st of July to the 2nd of October, a period which roughly covers the main spawning peak [25]. In the simulations, a constant number of larvae were released every 30 minutes and their movement followed during 60 days, which corresponds to the pelagic larval duration according to life cycle studies [16,26]. Thus, a total of 4465 larval release events were followed each year and their diffusive and advective movements averaged to arrive at a global dispersal kernel.

*P. pollicipes* larvae go through six naupliar stages which are adapted for dispersal, and one cyprid stage which does not feed and specializes in settlement [27]. There is evidence of ontogenetic vertical migration in some species of decapods [28] and in other stalked barnacles, with nauplii and cyprids occupying the surface and bottom layers, respectively (i.e. *Pollicipes polymerus*) [23]. Moreover, *P. pollicipes* nauplii exhibit both positive phototaxis and negative geotaxis (Gonzalo Macho, personal communication). Accordingly, in our simulations we have allowed the larvae to spend their six naupliar stages (approximately 30 days) [29] at the surface layer (0–10 m) and to sink and remain at the bottom layer (10–20 m) until completion of the planktonic phase as cyprids (Fig. 2).

## Physical background for the dispersal model

The larval dispersal process was simulated by means of a Gaussian probability density function, defined as

$$D_{ij} = (1/(L_S\sqrt{2\pi})) \exp((d_{ij} - L_A)^2/(2L_S^2))$$

where $D_{ij}$ is the probability density that a larvae released at point i reaches point j, $d_{ij}$ is the distance between points i and j, $L_S$ is the stochastic length scale and $L_A$ is the advective length scale [20]. The advective length scale refers to the net displacement of the larval population along the coast and across the shelf due to the directional (advective) component of the flow. The diffusive or stochastic distance indicates the extent to which the bulk of larvae have been spread around their average position by the turbulent component of the flow, given a certain decorrelation time scale (12 h at the coast) [19]. Separate simulations were run for the alongshore (X) and the cross-shore (Y) components of the flow, and the resulting probability fields were then combined to generate a 2D dispersal kernel. The probability function on the X-Y plane was calculated as the product of the proportion density functions on X and Y assuming they were uncorrelated [30]. This assumption was checked by means of linear regression analyses between currents on both axes for all depths during the 2009 and 2011 larval seasons ($R^2 \approx 0$ in all the cases). The total area under the normal curve must equal 1, i.e.

$$\sum (D_{ij}d_x) = 1 \quad \text{or} \quad \sum (D_{ij}d_y) = 1$$

being $dx$ and $dy$ distances on X and Y, respectively.

In the across direction, the coast is considered a "sticky boundary", that is, a domain where larvae are retained by means of active settlement behavior and/or reduced water motion [31]. Thus, inland (south of the shoreline) probability densities $D_{ij}$ (i.e., Y<0) were removed and added to those at the coast to preserve the $\sum (D_{ij}d_y) = 1$ condition.

*In situ* current measurements can be used to estimate the advective and diffusive length scales of the dispersal process. Our larval dispersal models were fed with *in situ* current velocities measured every 30 minutes in 10 two-meter depth cells using a Nortek Aquadopp Acoustic Doppler Current Profiler (ADCP) moored at 400 m off the Cudillero coast, N. Spain (43° 34.18′N 6°8.43′W; Fig. 1) at a 20 m depth. ADCPs do not capture the entire water column; therefore, we extrapolated the nearest measurements (2–4 m depth cell) to the surface cell (0–2 m) [32]. We then calculated the mean alongshore and cross-shore currents experienced by *P. pollicipes* larvae and their standard deviations for every possible spawning event during the 2009 and 2011 spawning seasons, that is, one event every 30 minutes taking into account their position in the water column (see preceding section). The currents experienced by each group of larvae during their pelagic larval duration (30 days in the surface layer followed by 30 days in the bottom layer) were averaged to obtain a mean advective length scale ($L_A$) and its standard deviation was used to calculate the stochastic ($L_S$) length scale [20].

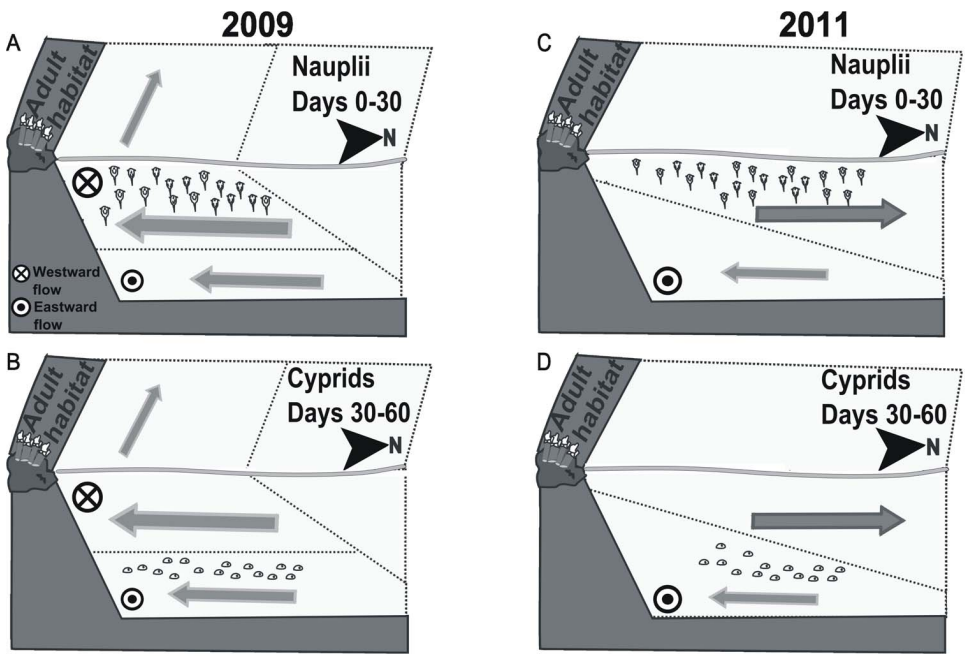

**Figure 2. Schematic representation of gooseneck barnacle larvae transport.** Nauplii (days 0 to 30, surface layer) and cyprids (days 30 to 60, bottom layer) are transported by the prevailing nearshore currents during high (left) and low (right) upwelling periods. Arrow thickness and symbol sizes are proportional to the average current velocities obtained from actual measurements (see figure 3). During day 0, stage I nauplii are released from the adult habitat to the surface layer. At day 30, stage VI nauplii turn into cyprids and experience an ontogenic vertical migration to the bottom water layer. At the end of their pelagic life at day 60, competent cyprids are transported to the adult habitat by southbound currents.
doi:10.1371/journal.pone.0078482.g002

Dispersal kernels for 2009 and 2011 were determined by ensemble averaging for all events in search of inter-annual differences in spatial dispersal patterns. In addition, to observe the general flow structure, alongshore and cross-shore velocity profiles were generated for both years using currents averaged over the entire larval season (1 July-31 November) for every 2 m depth cell (Fig. 3). Calculations and data processing were done with R 2.15.3 (R Development Core Team 2012), using the package *ggplot2* (Wickham 2009) for plotting.

## Characterization of the upwelling activity

Ekman transport in this region was characterized using two indexes: a daily upwelling index (DUI) and an integrated upwelling index (IUI). DUI is the average volume of water displaced per second and kilometer of coastline ($m^3 s^{-1} km^{-1}$). To calculate the DUI we have followed Bakun [33] as in Llope et al. [34] for the Cantabrian Coast, by using wind data collected at the Asturias airport meteorological station (43°33′N, 06°01′W, 127 m above sea level; Fig. 1) between 1969 and 2011. Wind intensities below 7 km h$^{-1}$ were set to 0, since these velocities were below the detection of early sensors. Positive upwelling index indicates that the surface layer is displaced off the coast and replaced by deeper water (i.e. upwelling process) while negative values indicate the reverse (i.e. downwelling). To summarize seasonal oscillations in the upwelling activity DUI data were fitted to a Generalized Additive Model (GAM)

$$DUI_t = f(t) + \varepsilon_t$$

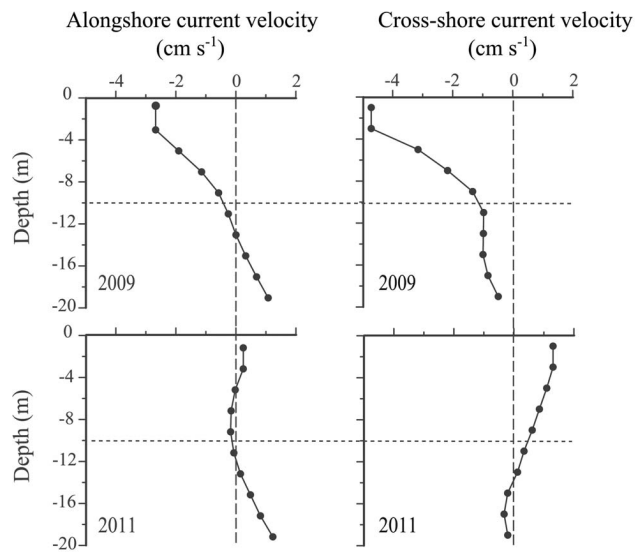

**Figure 3. Mean alongshore and cross-shore *in situ* current velocities.** Current velocity measurements for the gooseneck barnacle main spawning season (July-November) during a year of high (2009) and low (2011) upwelling. Vertical dashed lines separate westward (negative) from eastward (positive) alongshore currents and southward (negative) from northward (positive) cross-shore currents. The horizontal line separates the two water layers considered in the models.
doi:10.1371/journal.pone.0078482.g003

Where *t* is the day of year, *f* represents the smooth term (a thin plate regression spline), and $\varepsilon_t$ is the noise term [35]. To make the annual cycles comparable, we fixed the effective degrees of freedom of the model at 4, resulting in relatively smooth curves that give the general seasonal pattern. Model fitting was done using the *mgcv* package for R [36,37]. The IUI (total net volume of water displaced per kilometer of coastline per gooseneck barnacle recruitment season) was obtained by adding all daily cross-shelf Ekman transport measurements from June to October.

## Gooseneck barnacle catch rates

Monthly gooseneck barnacle landings and effort (days per fisherman) data collected during the fishing season (October-April) by the fishermen within each of 6 Asturian co-management plans were obtained from 1998 to 2011 (Fig. 1). The Luarca co-management plan dataset was incomplete and therefore had to be excluded from the analysis. Total catch rates for all 6 plans were determined as the sum of the landings divided by the sum of the effort in each fishing season. In the 2004–2005 fishing season a 2 kg reduction in the Total Allowable Catch (TAC) per fisherman was decreed. Thus, we modeled the effect of TAC change and IUI on catch rates using multiple linear regression models where data pre and post TAC change were identified with a dummy variable.

The estimated time between settlement and commercial size in gooseneck barnacles (21.50 mm rostro-carinal length) [38] ranges between 1 [26] and 5 years [39]. Thus, models were generated with lags between 1–5 years. A Bonferroni correction was applied to avoid excessive Type I error [40]. To ensure the accuracy of the coefficients, we tested that all the assumptions of a linear regression were met. Additionally, a model selection was performed using the adjusted R$^2$, the Akaike Information Criterion (AIC) and the Akaike weights. The variance explained purely by the IUI was determined by variance partitioning analysis following the approach by Legendre and Legendre [41]. Analyses were done using the *stats* package in R.

## Gene flow estimation and model comparison

*P. pollicipes* mitochondrial cytochrome c oxidase subunit I (COI) sequences from 5 locations of the Cantabrian coast were obtained from a previous study of Campo et al. [18]. Overall, we used 243 sequences coming from 2 populations of the Basque Country (Jaizkibel and Monpas); 2 populations of Asturias (Ribadesella and Punta de la Cruz) and 1 population from Galicia (Corme) (Fig. 4). To estimate migration between populations we applied a Bayesian approach to the COI sequences using software MIGRATE-N v3.216 [42,43]. Migration rates were estimated as the effective female population size in each sampling site ($N_{ef}$) multiplied by the migration rate (m) from and towards each site. Effective number of females in each population was estimated by the software, computing the Watterson estimator ($\theta = N_{ef} \cdot \mu$) using coalescent equations [44], which does not need *a priori* knowledge of the mutation rate of the locus ($\mu$). Migration rate from and towards each site was derived from its mutation-scaled migration rate value ($M = m \cdot \mu$).

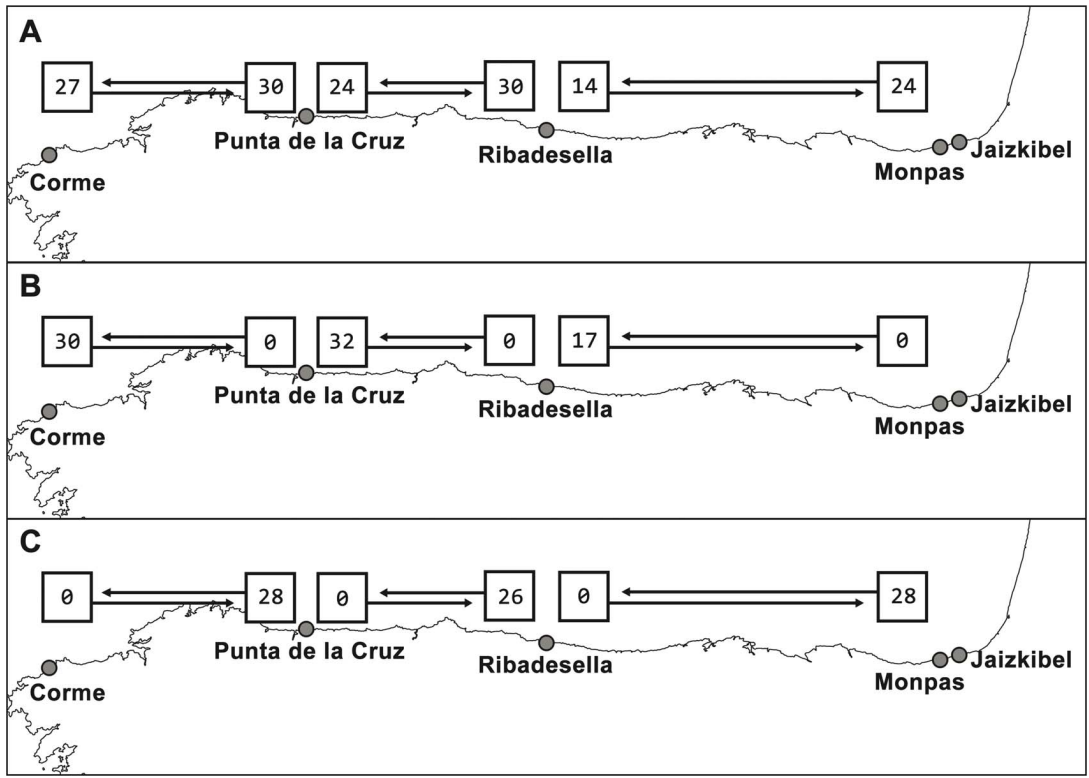

**Figure 4. Schematic diagram of the MIGRATE model results superimposed on a map of the Cantabrian Coast.** The map shows the effective number of migrants per generation ($N_{ef} \cdot m$, numbers within squares) and the direction of migration (arrows). A: Full model; B: Eastward model, westward dispersal set to zero; C: Westward model, eastward dispersal set to zero.
doi:10.1371/journal.pone.0078482.g004

Analyses were first run with a full migration matrix in which gene flow was unrestricted between neighboring populations. To explicitly test for eastward or westward larval dispersal, we also tested two custom unidirectional matrices in which gene flow was only allowed in one direction in every population. All computations on these matrices were performed with two long Markov chains of 10 million generations and a static four-chain scheme with default heating values. Default uniform priors and slice samplers were used for the M and $\theta$ parameters, and starting run values were estimated from the $F_{ST}$ measure as computed by the software.

Finally, likelihood scores for all migration models were obtained by a thermodynamic integration with Bezier approximation [45], as implemented in the software. Direct comparison of models was then assessed by manually transforming these likelihood scores into Bayes Factors [46] which was performed using the method described in Beerli and Palczewski [47].

## Results

### Upwelling regime in the Central Cantabrian Coast

The fitted DUI general cycle in the Cantabrian coast for the last 40 years is a unimodal curve with positive values during the gooseneck barnacle recruitment season (p-value <0.0001; Fig. 5). This process has been observed by other authors who have stated the appearance of upwelling processes throughout the summer period in the Cantabrian Sea [12,48]. The values for 2009 follow the general pattern (p-value <0.0001); in contrast, 2011 presents a bimodal shape with positive values both at the beginning and the end of the recruitment season, but negative in the middle (p-value <0.0001) (Fig. 5). This difference is also apparent in the sea surface temperature, which was approximately 3°C higher in 2011 than in 2009 (Fig. 1). In 2011, upwelled water masses and their characteristic onshore flow were restricted to the bottom water layer (Fig. 2, Fig. 3) and offshore exportation at the upper layer of the naupliar stages may occur. In contrast, upwelled waters moving westwards and shorewards spanned almost the entire water column in 2009 (Fig. 2, Fig. 3) due to the intense upwelling activity registered (Fig. 1, Fig. 3).

### Simulations of larval dispersal

Simulated net mean displacement of the larval population for 2009 was westwards (alongshore $L_A = -56.12 \pm 21.8$ km, mean±SD) and landwards (cross-shore $L_A = -67.79 \pm 47.3$ km), while in 2011 it was slightly eastwards (alongshore $L_A = 12.95 \pm 9.8$ km) and seawards (cross-shore $L_A = 31.61 \pm 8.4$ km). 2011 experienced lower flow variability than 2009, with alongshore $L_S$ ($24.294 \pm 2.15$, $42.790 \pm 7.77$, respectively) and cross-shore $L_S$ ($28.478 \pm 3.57$, $46.645 \pm 9.17$), leading to less dispersed larval distributions (Fig. 6). Examples of the modeled larval dispersal events can be viewed in the animations (time-series of raster images) for the first event of each season (July 1st to August 29th 2009 and 2011; see Animations S1 and S2).

High probability densities ($D_{ij}$) were obtained far from the coast in 2011 ($22.88 \times 10^{-5}$ km$^{-2}$), which entails potential larval losses to the adult population (Fig. 6). However, due to the "sticky boundary" condition imposed to the shoreline in our models (see methods) $D_{ij}$ maximum values ($49.9 \times 10^{-5}$ km$^{-2}$) were reached at the coast (Fig. 6). In 2009 the probability density function was constrained to the coast and displaced 56 km westwards of the release point. Maximum $D_{ij}$ values for both breeding seasons were reached at the shore indicating potential recruitment to the adult populations. However, $D_{ij}$ values at the coast in 2009 were 3 times higher than in 2011 ($173.8 \times 10^{-5}$ km$^{-2}$ compared with $49.9 \times 10^{-5}$ km$^{-2}$, Fig. 6), entailing a greater theoretical recruitment success in 2009 (94%) than 2011 (15.4%). We tested the hypotheses of our models, westward bias and increased recruitment during upwelling years, through gene flow patterns and catch rate data.

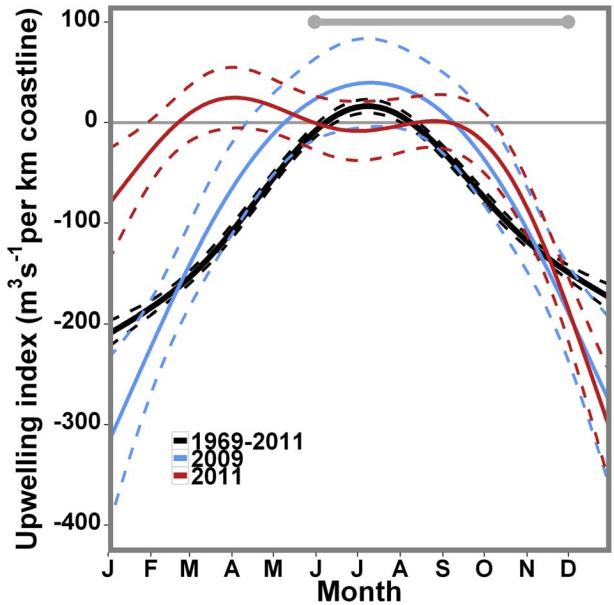

**Figure 5. Fitted, Daily Upwelling Index vs. Julian Day GAM regression lines.** Regression lines represent the whole wind data series (1969 to 2011, black) and years of high (2009, blue) and low upwelling activity (2011, red). Dotted lines depict the associated standard error. The grey horizontal line covers the gooseneck barnacle recruitment season.
doi:10.1371/journal.pone.0078482.g005

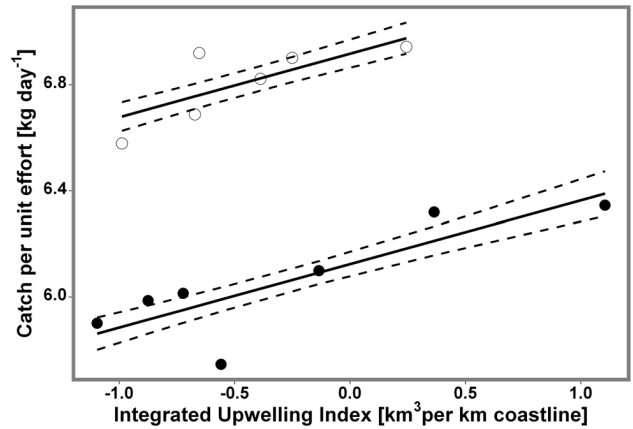

**Figure 7. Relationship between catch rates per fishing season and the Integrated Upwelling Index 4 years before.** Dots represent years before the change in TAC (hollow) and after (shaded). Dashed lines indicate the standard error.
doi:10.1371/journal.pone.0078482.g007

Migration rates for Cantabrian gooseneck barnacle populations were estimated to test the alongshore component obtained from our simulations. Gene flow patterns inferred by MIGRATE are shown in Figure 4 for the three computed models. Migration rates, being far higher than one individual per generation, are indicative of a high historical connectivity between all populations [49], but values estimated in the full model (A) suggest a dispersal pattern biased towards the west. Comparison of all models indicate that the full model is the most appropriate for our dataset followed by the westward and eastward models according to their likelihood ($-1945.63$, $-1984.14$, $-2009.86$, respectively), the natural logarithm of their Bayes factors (0, $-77.02$, $-128.47$) and their resultant Bezier probability ($\approx 1$, $3.54 \times 10^{-34}$, $1.612 \times 10^{-54}$).

The effect of upwelling intensity on population recruitment (i.e. cross-shore component in our simulation) was inferred through catch rate data. Using Akaike Information Criteria (AIC) and adjusted $R^2$ for model selection, we determined that the 4 year lag model best describes the relationship between IUI and catch rates using TAC as a dummy variable (Table 1). The model explains 94% of the variability in catch rates. After applying a

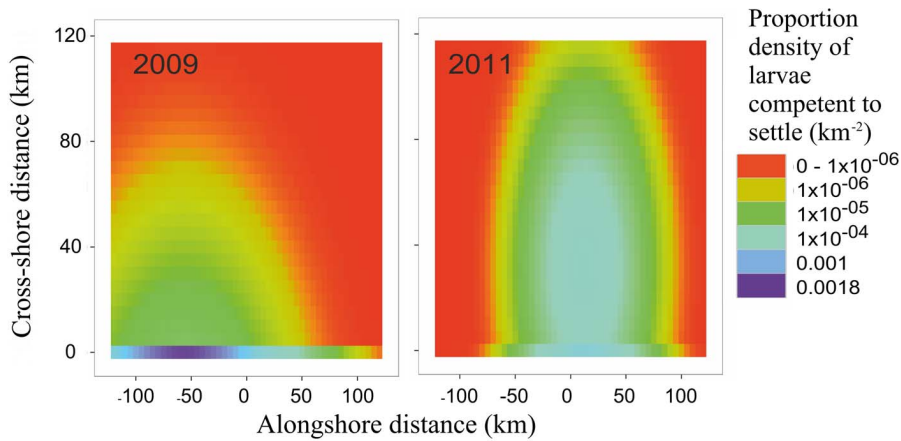

**Figure 6. Distribution kernels for *P. pollicipes* larvae.** Distribution of the proportion of gooseneck barnacle larvae (km$^{-2} \times 10^5$) released at the origin of coordinates which are competent to settle after completing their pelagic larval duration (60 days, see methods). Larvae not located at the coast are considered lost to the population. Each pixel has 5 km$^2$.
doi:10.1371/journal.pone.0078482.g006

**Table 1.** Comparisons for the model (catch rates$_i$ = α + βIUI$_i$ + TAC$_i$ + $\mathbf{E}_i$) using 1–5 year lags in IUI.

| Model | p-value | Adj. R² | AIC | AIC weight |
|-------|---------|---------|-----|-----------|
| **1 year lag** | 0.0002 | 0.7891 | −0.7664 | 0.0007 |
| **2 year lag** | 0.0001 | 0.8058 | −1.8366 | 0.0012 |
| **3 year lag** | 0.0001 | 0.7964 | −1.2270 | 0.0009 |
| **4 year lag** | <0.0001 | 0.9308 | −15.2442 | 0.9961 |
| **5 year lag** | 0.0001 | 0.8011 | −1.5263 | 0.0010 |

Bonferroni correction both explanatory variables TAC (p-value = <0.00001) and IUI (p-value = 0.005), remained sigñificant. IUI displayed a significant positive correlation with catch rates (Fig. 7). The results from our model were

$$\text{Catch rates} = \begin{cases} 0.24 \times IUI + 6.125 & \text{if } TAC = 6kg \\ 0.24 \times IUI + 6.918 & \text{if } TAC = 8kg \end{cases}$$

This corresponds with an increase in catch rates of 0.24 kg day$^{-1}$ for every km$^3$ displaced per km of coastline during the recruitment season 4 years before. According to the variance partitioning analysis, TAC explains 90% and IUI 12% of the variance in catch rates.

## Discussion

### Asymmetric connectivity patterns within gooseneck barnacle populations

Our genetically inferred migration rates (Fig. 4) indicate a high level of gene flow among Cantabrian Sea *P. pollicipes* populations; a conclusion consistent with those of Campo et al. [18], using phylogenetic methods on the same dataset, and Quinteiro et al. [17], based on samples collected along the full geographical range of the species. Although any inferences of gene flow based on a single locus should be taken with caution [50], the absence of recombination and high mutability of mitochondrial loci, such as COI, makes them useful for inferring patterns at local scales [51], especially when testing the fit of particular models [52], as is our case.

According to Quinteiro et al. [17], genetic patterning of *P. pollicipes* is a consequence of long range larval dispersal driven by the IPC, a slope current which circulates in an eastwards direction in the Cantabrian coast. However, this hypothesis is at odds with the result of our migration rate analysis, which points to the prevalence of a westwards flow. Certainly, the IPC is a major structuring agent of planktonic populations in the Central Cantabrian Sea [34,53,54]. It is a seasonal structure characteristic of late autumn and winter that reaches its maximum extension by the end of it year [55,56]. However, gooseneck barnacles reproduce and release their larvae mainly between April and September [16,24], a period when the IPC is at its lowest and coastal upwelling at its highest [13,57]. Furthermore, the IPC is usually far from the thin strip of nearshore water where the larvae are released and recruited. Thus, it is unlikely that the IPC plays a major role in larval dispersal of this species, which rather depends on summer hydrography, and more specifically on the activity of the NW Spanish Upwelling system, which in this area is highly variable both within and between years [57].

The westward-biased connectivity pattern, which is even seen in the unrestricted migration rate model, suggests that the structure observed in the larval dispersal kernel persists in the adult population. Thus, the recurrence of upwelling may not only define the spatial scale and direction of the dispersal process but also the genetic structure of the barnacle metapopulation. Such an effect of environmental oceanic conditions on genetic patterning has been observed in other species with a pelagic dispersal stage [58,59,60], and while its influence does not seem remarkable in mesoscale or large-scale space [61], it might have also been underestimated in local processes of population differentiation and patterning for several species [62].

The asymmetrical connectivity pattern observed in our results is consistent with recent findings by Nolasco et al. [63] and Domingues et al. [64] in the western Iberian Peninsula. These authors employed a biophysical simulation to analyze connectivity among populations of *Carcinus maenas*. Their results indicate a southward bias of approximately 61 km in a 9-year average dispersal kernel. Larval dispersal patterns are attributed to the effect of summer upwelling events in the area. Our results for a year of high upwelling activity (2009) match both the magnitude (56 km) and asymmetry in larval dispersal (Fig. 6) determined by these authors; providing further evidence against IPC-mediated connectivity patterns for gooseneck barnacle populations in the Iberian Peninsula.

### Upwelling and recruitment

Active coastal upwelling during the gooseneck barnacle recruitment season has been the most frequent situation during the last 40 years in the Cantabrian Sea (Fig. 5). Other authors have stated the appearance of upwelling processes throughout the summer period in the Cantabrian Sea [12,48]. Typical summer upwelling circulation in the region is characterized by a westward flow driven by northeasterly winds [65]. This is consistent with the direction of alongshore dispersal calculated with the ADCP data for both years (Fig. 6). However, although they agree with the expected oceanographic patterns, it must be taken into account that the dispersal kernels are based on current measurements for only 2 years, 2009 and 2011, at a single location. Coastal dynamics are highly variable in our system [34]; hence, a single mooring 400 m away from the coast might not be able to accurately represent the nearshore dynamics for the entire coast.

According to the classical paradigm of upwelling circulation, strong advective surface flow off the coast associated with upwelling activity leads to larval export towards the open ocean. In this scheme, it is only through relaxation of the upwelling that the exported larvae have any chances of recruiting to the coastline by reversion of the surface water flow [66,67]. In contrast, our biophysical simulation predicts increased recruitment during strong upwelling years (Fig. 6). *In situ* current measurements for 2009 (intense upwelling year) reveal an onshore return flow across the water column (Fig. 3) which should promote a nearshore retention of larvae. This type of flow during intense upwelling events has been previously documented in the area [68]. In fact, this onshore return flow is common in some upwelling systems at narrow shelves and sloping bottom profiles [69], such as the Cantabrian Sea. Our results agree with recent findings in the western Iberian Peninsula [63] and northern California [70], where the interaction between upwelling activity and active vertical swimming behavior favors recruitment by retaining larvae close to the coast.

Furthermore, Pavón [16] observed a strong correlation between northeasterly winds, responsible for summer upwelling episodes in the Cantabrian Sea, and recruitment of *P. pollicipes* in 2 locations at the western Asturian coast. Likewise, a correlation between recruitment and upwelling has been observed for *P. pollicipes* in Cabo Sines, Portugal [39] and for *P. polymerus* in the Southern

California Bight [71]. Thus, our results point to the importance of the vertical distribution of barnacle larvae and of its interactions with the nearshore hydrography as a determinant of their recruitment.

Our results also indicate that the effect of upwelling on gooseneck barnacle recruitment will have consequences on the productivity of this species in the area. Most of the variability found in Asturian gooseneck barnacle catch rates is explained by changes in the total allowable daily catch, which changed in 2005 from 8 to 6 kg per day per fisherman (Fig. 7). However, all the remaining variability (12%) is explained by the IUI, which suggests that fluctuations in recruitment rates predicted by our biophysical model and those observed by Pavón [16] in the Asturian Coast translate to variations in the adult population. In our optimal model, a time lag of 4 years was allowed between the IUI (i.e. recruitment of the larvae) and the catch rate series (Table 1). Unlike *P. polymerus*, whose age at sexual maturity and commercial size are known (1 and 5 years, respectively) [72], estimates of age at commercial size in *P. pollicipes* are highly variable [16,26,38,39], with an average of 3.4 years, when considering their different growth rates estimates. This is fairly consistent with the time lag of the best scored of our alternative models (Table 1). Therefore, a year of active upwelling should enhance the production of catchable *P. pollicipes* 4 years afterwards.

## Management implications

Sustainable management of gooseneck barnacle fisheries in the western Cantabric coast is carried out through individual daily quotas and partial closures of groups of rocks [10,16]. The timing and exact location of each closure is decided each year through consensus among the administration and the fishermen belonging to each of the 7 co-management plans, demonstrating a model example of adaptative management. Total bans are spaced heterogeneously from 0.2–20 km with a length between 0.1 and 3 km for each area. These bans are implemented exclusively in overharvested or economically important areas to prevent over exploitation (Fig. 1). This strategy avoids a decrease in the exploitable stock and protects from population "washout" [73] in the target area, thanks to the effect of the diffusive component of the flow which allows a small proportion of larvae to settle at the emission point (Fig. 6).

Despite the continuous efforts to prevent the overexploitation of gooseneck barnacle fisheries, aspects such as larval dispersal scale and direction have not been taken into account in the co-management system. Our results suggest that when developing management guidelines concerning the location and distribution of bans, a geographically biased dispersal pattern should be acknowledged. Protection of specific target areas by means of closures should be complemented with bans in rocks located to the east of those areas. Likewise, to determine areas with a high yield, catch rates for the fishing season can be estimated using the TAC for the year and the IUI in the recruitment season 4 years before.

Furthermore, in our biophysical simulations typical *P. pollicipes* dispersal distances ranged between 10 and 60 km (Fig. 6), thus local conservation strategies have the potential to interact with population persistence. The Asturian gooseneck barnacle management plan is an ideal place to develop these strategies considering the active co-management system in the area and their rotating rock closure strategy (Fig. 1). A network of total bans can be established in the co-management system by retaining their current size but redistributing the bans from clusters to evenly spaced rock closures at the gooseneck barnacle dispersal scales. These areas can act as temporal small-scale marine protected areas where larvae can disperse among reserves, ensuring the persistence of the population. A similar reserve distribution has been suggested by Hastings and Botsford [74], for species with a pelagic larval stage and sessile adults, as the optimal reserve arrangement to achieve an increased fisheries yield and at the same time ensuring the sustainability of populations. This is also consistent with findings in Gaines et al. [75], which indicate that multiple reserves are more effective than single reserves of the same total size in areas strongly affected by currents, such as the Cantabrian Sea.

Our results reveal a clear role of upwelling on *P. pollicipes* larval dispersal and population connectivity in the Cantabrian Sea. However, considering the inherent variability in the NW Spanish Upwelling system, continuously changing management guidelines need to be employed to incorporate such effect. In this regard, the current adaptative character of the Asturian co-management system favors the incorporation of these measures.

## Conclusions

Our results reveal a clear imprint of upwelling on the genetic structure and productivity patterns of gooseneck barnacle meta-populations. In spite of being produced by a hydrographic structure on a scale of a few hundred kilometers, the scale of these effects (10–60 km) is perfectly consistent with the management units. Such effects should therefore be incorporated in a sound management strategy. In this paper we have suggested possible management measures, according to the dispersal scales and connectivity patterns obtained through our biophysical simulations, which could be advisable to incorporate in the local gooseneck barnacle co-management system.

## Supporting Information

**Animation S1 Daily time-series of the first larval dispersal event in the Asturian coast for 2009.** In the first event considered in our model (see methods) larvae are released on July 1st, they finish their naupliar stage and turn into cyprids on July 31st and finalize their pelagic larval duration on August 29th. The cross indicates the emission and maximum recruitment points. Users can view the daily progression of $D_{ij}$ values for these points by clicking on the cross.
(KMZ)

**Animation S2 Daily time-series of the first larval dispersal event in the Asturian coast for 2011.** In the first event considered in our model (see methods) larvae are released on July 1st, they finish their naupliar stage and turn into cyprids on July 31st and finalize their pelagic larval duration on August 29th. The cross indicates the emission and maximum recruitment points. Users can view the daily progression of $D_{ij}$ values for these points by clicking on the cross.
(KMZ)

## Acknowledgments

We thank the Agencia Estatal de Metereologia (AEMET) for providing wind data and the Centro de Experimentación Pesquera del Principado de Asturias (CEP) for their logistic support, particularly Jorge Alcázar and Alberto Prada. We also thank the NASA Ocean Biology Processing Group at the Goddard Space Flight Center and Information Service Center (Giovanni service) for the production and availability of MODIS data. Carlos Cáceres, Daniel Campo and Eva García provided valuable comments for this manuscript. We greatly appreciate the genetic sequence data provided by Daniel Campo and Eva García as well as the information provided by Gonzalo Macho.

## Author Contributions

Conceived and designed the experiments: AR NW AFP. Performed the experiments: AR NW AFP RGG. Analyzed the data: AR NW AFP RGG.